# Atomic layer deposition of aluminum (111) thin film by dimethylethylaminealane precursor

Sameh Okasha[*a&b], Yoshiaki Sekine[c], Satoshi Sasaki[c], and Yuichi Harada[a]

[a] Global Innovation Center, Kyushu University, 6-1 Kasuga-Koen, Kasuga, Fukuoka 816-8580, Japan

[b] Molecular and material science department, Interdisciplinary Graduate School of Engineering Sciences, Kyushu University, Kasuga, Fukuoka 816-8580, Japan

[c] NTT Basic Research Laboratories, NTT Corporation, 3-1 Morinosato Wakamiya, Atsugi, Kanagawa 243-0198, Japan

**Abstract:**

We report the growth of aluminum (111) thin film by atomic layer deposition (ALD) technique with dimethylethylaminealane (DMEAA) as a precursor. It is found that the metallic underlayer is essential to grow uniform aluminum films by DMEAA precursor. As a titanium thin film is used as the underlayer, grown aluminum thin film shows (111) orientation irrespective of substrates. The lattice constant and superconducting transition temperature of the aluminum thin films are the same as the bulk one. These findings suggest that ALD technique provides high quality of the aluminum thin films and have potential for the applications of superconducting devices. We discuss ALD technique with DMEAA precursor is the promising method for fabricating vertical small Josephson tunnel junctions, which can be used as the superconducting quantum bits.

[*] Corresponding author: sokasha@kyudai.jp

1.  **INTRODUCTION**

Atomic layer deposition (ALD) technique enables us to fabricate atomically controlled thin films [1]. ALD technique can deposit insulating materials conformally which can cover the whole surface of nanowires [2] and realize quantum point contact made of a narrow $In_{0.75}Ga_{0.25}As$ channel [3]. It can be also used as the passivation layer in order to stabilize adsorbed molecules [4]. ALD technique can also deposit metallic thin films and show a sharp interface between different materials as the sequential deposition [5]. The rather low growth temperature of ALD technique offers in situ photoresist processes. These features are promising to fabricate stacked tunnel junctions including a Josephson tunnel junction that is composed of two superconducting electrodes separated by a very thin (~ 2 nm) insulating barrier.

There are several reports on precursors for the ALD growth of aluminum thin films. Trimethylaluminum (TMA) and water is a standard process to grow aluminum oxide [1], while TMA and hydrogen is used to grow aluminum thin film[6]. Dimethylaluminum hydride (DMAH) is also a good candidate as the precursor which offers reduced carbon contamination [7] and an area-selective growth[8]. Among the three precursors, we decide to use dimethylethylaminealane (DMEAA), which has several advantages: a low decomposition temperature [9], avoiding carbon contamination with long-term sustainability [10]. Although some groups reported the superiority of DMAH compared with DMEAA [7], DMAH has the disadvantages of the rough surface.

Here, we study the growth of aluminum thin films by ALD with DMEAA as a precursor. We found the metallic underlayer is necessary for obtaining aluminum. The ALD-grown aluminum films show superconductivity as same as bulk aluminum.

## 2. Experimental details

Aluminum thin films are grown in the thermal ALD chamber (SUNALE R-150 by Picosun oy.) equipped with a glovebox where nitrogen gas is filled up in order to prevent samples from oxygen exposure. DMEAA [11,12] precursor is used as the source of aluminum. It is worth noting that the viscosity of DMEAA is stable so that the reproducibility of deposition conditions for aluminum thin films is high while we found that the viscosity of DMAH changes in six weeks and thereby it is very difficult to grow reproducibly aluminum thin films under the same conditions. We used silicon (Si) substrates for aluminum deposition. Two of Si-substrates are prepared, one as bare substrate and the latter with a metallic layer. Both substrates are prepared with the standard degrease process and surface etching by tetramethylammonium hydroxide solution [13]. It is found that aluminum thin films can be grown on the metallic layer [14]. As a metallic underlayer, a 1.5 nm-thick titanium film is deposited on the silicon substrate by electron-beam with a very low deposition rate of 0.05 nm/sec., which gives a smooth surface. After the deposition of Ti seed layer on the Si substrate, the substrate must be installed into the ALD chamber within half an hour in order to avoid the surface oxidation.

The growth of aluminum from DMEAA is not a self-limited process, thus an aluminum monolayer (ML) growth can be tuned by the combination of an amount of DMEAA which is determined by the valve opening time of DMEAA, $t_{open}$, purging time removing by-products from the chamber, $t_{purge}$, and the growth temperature in the chamber, T. The ALD chamber is equipped with the hot-wall and thus the whole chamber is kept at the same temperature during the film growth. By adjusting three conditions, we can obtain 0.3 ML to 4 ML growth per cycle (GPC). The following results were obtained with the growth conditions of 300 ALD cycles that $t_{open}$, $t_{purge}$, and T are 0.3 seconds, 3.0 seconds, and 150 degrees Celsius, respectively.

The surface morphology of one of the aluminum thin film and its thickness are characterized by field emission scanning electron microscopy (SEM, JEOL JSM-6340F capable of getting image resolution of 3 nm). The rough estimate of the surface roughness was obtained from a 45° tilted view of cross-section SEM image of an Al film prepared over a Ti underlayer. No signal filtering was applied and data was corrected for the tilted geometry. The average distance between the edges of the Ti layer and along the lumen was evaluated. Moreover, the maximum peak-to-valley (PV) distance and root mean square (RMS) roughness of the edge of the Al film with the lumen were also estimated, all of these with the help of the Gatan digital micrograph ver. 3.4. The composition and crystal orientation of the two films were estimated from X-ray diffraction (XRD) measurements performed with XRD (Rigaku TTR-III) apparatus, using a Cu Kα x-ray source (1.5406 Å) and the apparatus software package.

The resistance measurements were performed into the 0.3 K refrigerator with a 3 T superconducting magnet. The temperature was measured by the ruthenium oxide thermometer located at the sample stage.

3.  **Results and Discussion**

Figure 1 (a) and (b) show the SEM images of the grown thin film of aluminum on a silicon substrate without and with a titanium underlayer, respectively. We grow aluminum thin films onto the bare silicon substrates at various growth conditions. However, in the absence of a Ti underlayer, only aluminum nano-grains were obtained on the Si substrate as shown in Fig. 1 (a). On the other hand, as shown in Fig. 1 (b), aluminum thin film with the Ti underlayer shows a mirror-like smooth surface. We speculate that Ti removes hydrogen ions from alane ($AlH_3$) in a similar mechanism of the decomposition of alane by titanium (IV) isopropoxide [15]. Indeed, aluminum films can't be obtained when a 1.5 nm thick Ti underlayer is exposed into the ambient atmosphere for more than half an hour due to the surface oxidation of a Ti film. The average thickness and standard deviation of the aluminum film grown for 300 ALD cycles with a GPC of 3ML/cycle are 220 ± 33 nm, and estimates of the RMS roughness and maximum observed peak-to-valley of the cleaved edge of the sample are 16 nm and 150 nm, respectively. The material choice of the underlayer is also considered. We found gold sometimes produces intermetallic compounds with aluminum, which may become an insulator [16]. We choose a very thin titanium film as the underlayer in this work.

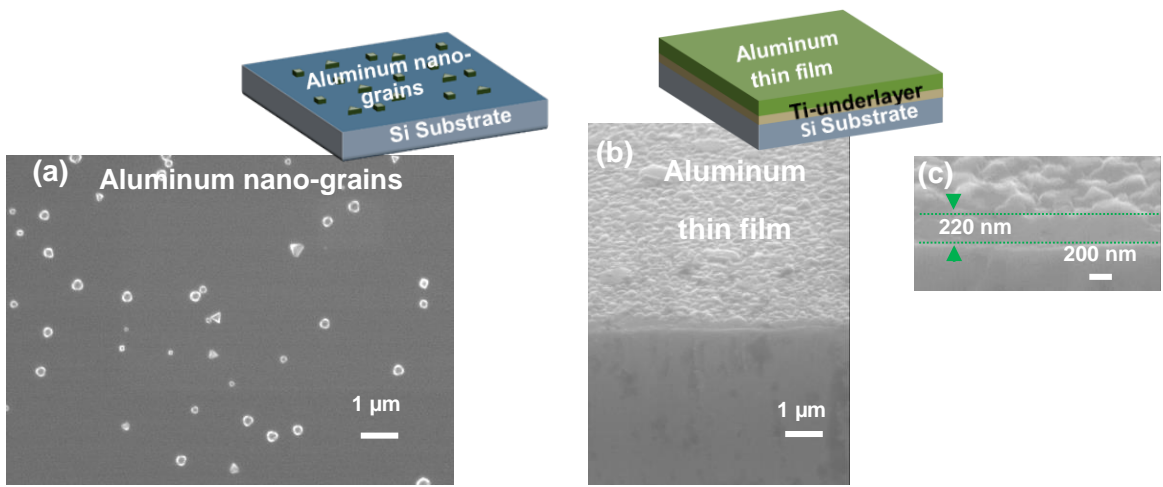

Fig. 1. SEM images with respective illustrative drawings of (a) Top view of aluminum nano-grains grown on the bare Si substrate, (b) tilted-view (45°) of the aluminum thin film on the Si substrate with a titanium underlayer, and (c) cross section image of average-thickness of aluminum thin film.

XRD measurements indicate that an aluminum film with the Ti seed layer on the Si substrate has (111) orientation with a peak of $2\Theta^o$ = 38.5 degrees [17] as shown in Fig. 2. The peak at $2\Theta^o$ = 69 degrees represents Si (400) orientation of the substrate. Al (111) is an ideal orientation due to the high resistance of stress migration with a longer lifetime as an electrode for device applications [17,18]. We also grew aluminum films on other substrates, such as GaAs and sapphire, with a Ti under layer and found all of the aluminum films show (111) orientation. Our results are coincident with the reference [17].

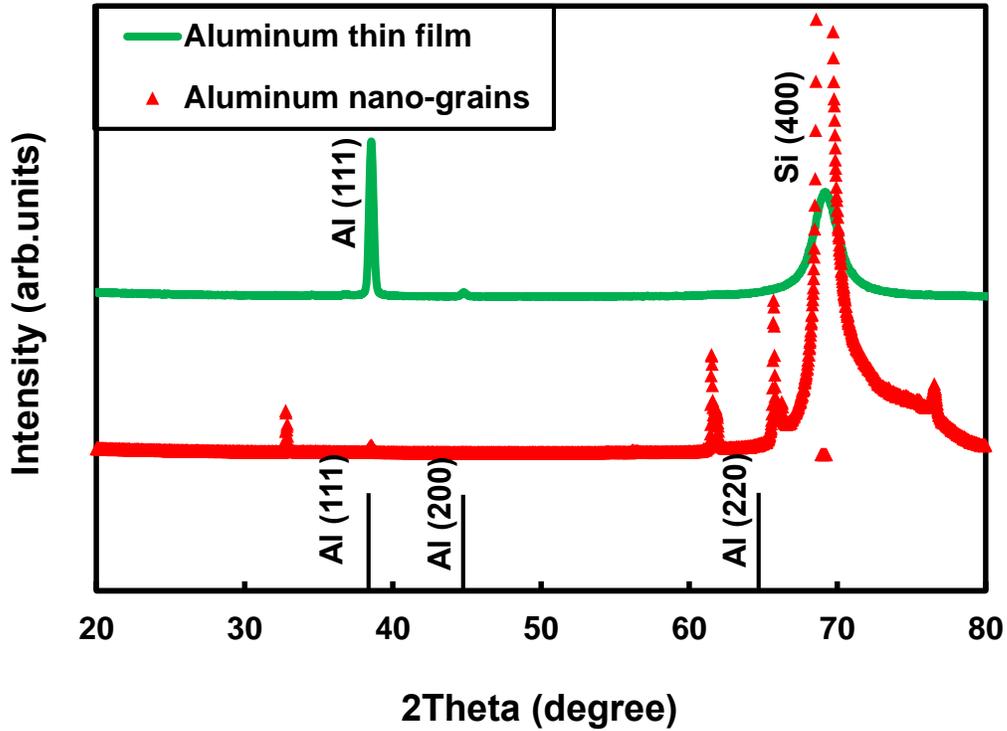

Fig. 2. XRD analysis shows the aluminum texture of (111) orientation at 2Θ º = 38.5º for both the thin film and nano-grains along with stacked lines of the standard reference at the bottom. The peak at 2Θ º = 69 degrees represents Si (400) orientation of the substrate.

Further analysis for the XRD peak of Al (111) at 2Θ º = 38.5 degrees is performed to acquire crystal information of the grown thin film. The full width at half maximum (FWHM) value was obtained to be 0.295 degrees. The plane distance of the grown aluminum (111) thin film was derived to be 0.233 nm. The mean grain size of the aluminum thin film was 32 nm calculated from the Scherrer equation as following: $Lc = (K\lambda)/(FWHM*Cos\theta)$, where Lc is crystallite size, K = 1.

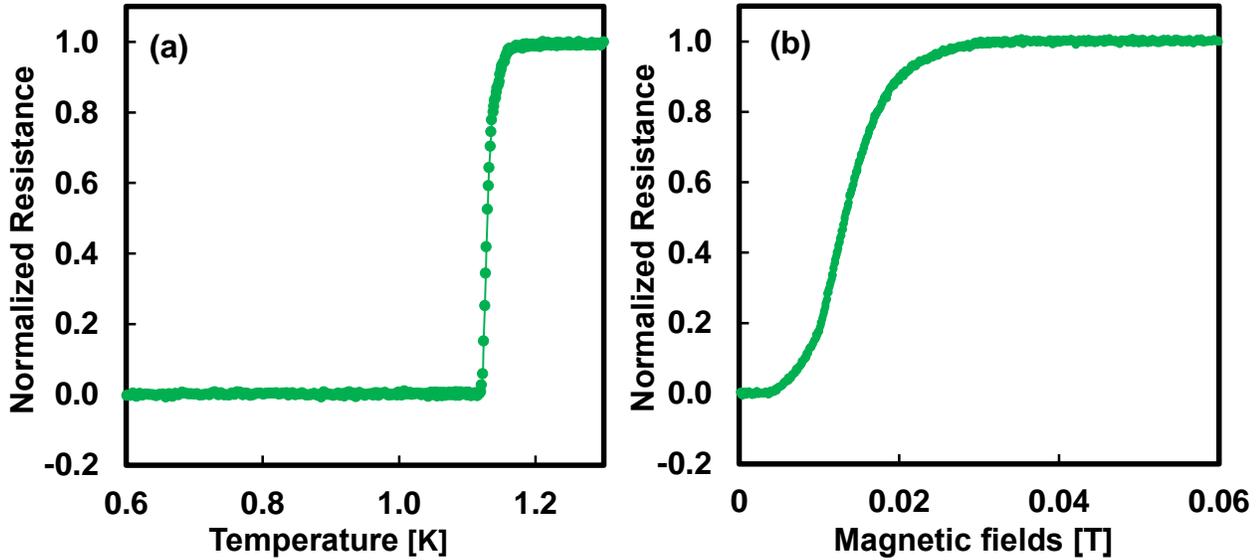

Fig. 3. (a) The temperature dependence with zero magnetic fields and (b) the magnetic field dependence with T= 0.3 K of the aluminum film with the thickness of 110 nm. The resistance is normalized.

Figure 3 shows the temperature and magnetic field dependence of the aluminum film with the thickness of 110 nm. The onset of the superconducting transition temperature is 1.18 K which is very close to the bulk value, 1.19 K [19]. It is well known that thin aluminum films show the enhancement of superconductivity due to either impurity effects [20] or phonon softening [21,22]. As measured by XRD, the (111) plane distance of 0.233 nm for the grown aluminum thin film Al (111) plane distance of the Al thin film is almost same as that of 0.234 nm for bulk aluminum [23]. These suggest that grown Al films are very pure and ideal thin films. However, the magnetic field dependence of the aluminum film at 0.3 K shows rather a broad transition although aluminum is a type I superconductor, which implies that the distribution of aluminum grains may play a role [24].

As we already pointed out, aluminum film only grows onto the metallic layer, not on the bare silicon substrate. This gives the selective growth of aluminum thin film onto the Ti seed layer and it is very useful for fabricating superconducting small Josephson tunnel junctions, which have the nominal area less than 100 x 100 nm$^2$. Current standard fabrication of small Josephson tunnel junctions uses the shadow evaporation technique with a suspended bridge [25]. As the etching process of small Josephson tunnel junctions causes the degraded junction quality, the suspended bridge technique is only the way to fabricate small Josephson tunnel junctions. However, the suspended bridge technique leaves unnecessary patterns due to the pattern shift [26] and it makes difficult to fabricate large integrated circuits of small Josephson tunnel junctions. Our findings of the selective area growth of aluminum thin films may enable us to fabricate the self-aligned vertical type of small Josephson tunnel junctions and contribute to a way of making superconducting quantum circuits, such as superconducting quantum bits.

## 4.   Conclusion

We grew uniform aluminum (111) thin film using the atomic layer deposition technique with dimethylethylaminealane precursor. A smooth aluminum thin film with strong texture of (111) is obtained by using the titanium underlayer regardless of the substrate materials. The superconducting transition temperature of the grown aluminum thin films is the same as the bulk one and the lattice constant of the thin films is coincident with the bulk one, which indicates aluminum thin films are high quality. The selective area growth has the potential to obtain the way to fabricate self-aligned vertical Josephson tunnel junctions.

ACKNOWLEDGMENTS

One of the authors is grateful for the financial assistance provided by the green Asia program at Kyushu University.